\documentstyle[12pt]{article}

\newcommand{\be}{\begin{equation}}   \newcommand{\ee}{\end{equation}}
\newcommand{\bear}{\begin{eqnarray}}
\newcommand{\eear}{\end{eqnarray}}
\newcommand{\ba}{\begin{array}}      \newcommand{\ea}{\end{array}}
\begin{document}
\pagestyle{empty}
\begin{titlepage}

\vspace*{-8mm}
\noindent 
\makebox[11.5cm][l]{BUHEP-96-7}  September 28, 1996\\

\vspace{2.cm}
\begin{center}
  {\LARGE {\bf Baryogenesis from Cosmic Strings  \\ [2mm] 
               at the Electroweak Scale. }}\\
\vspace{42pt}
Indranil Dasgupta \footnote{e-mail address:
dgupta@budoe.bu.edu}

\vspace*{0.8cm}

{\it \ \ Department of Physics, Boston University \\
590 Commonwealth Avenue, Boston, MA 02215, USA}

\vspace*{0.5cm}

\vskip 3.4cm
\end{center}
\baselineskip=18pt

\begin{abstract}

{\normalsize

{We explore the viability of baryogenesis from light scalar decays
{\it {after}} the electroweak phase transition. A minimal model of this
kind is constructed with new $CP$ violating interactions
involving a heavy fourth family. 
The departure from thermal equilbrium must come from
topological defects like cosmic strings, and we show that almost any
mechanism for producing the cosmic strings at the electroweak scale
results in a viable theory. 
Baryogenesis occurs in the fourth generation but the baryon number
is later transported to the visible generations. This mechanism of
{\it {indirect}}
baryogenesis allows us to satisfy experimental limits on the proton
lifetime while still having perturbative baryon number violation at low
energies. The fourth family has very small mixing angles which opens the
possibility of 
distinct observable signatures in collider experiments.}}

\end{abstract}

\vfill
\end{titlepage}

\baselineskip=18pt
\pagestyle{plain}
\setcounter{page}{1}





\section {Introduction}

The experimental bounds on the ratio of baryon excess to
the entropy of the universe are \cite {kolbturner}:
\be \label {introbar} 
\eta = (2 - 8) \times 10^{-11}.
\ee
Baryogenesis is an attractive explanation
of the observed fact that baryons are more abundant than anti-baryons in
the universe. 
The conditions necessary for baryogenesis were spelled out by Sakharov
nearly three decades ago \cite {sakharov}. The three conditions are: (i) existence of baryon number
violating interactions; (ii) C and $CP$ violating processes; and (iii) a departure from
thermal equilibrium. The earliest models of baryogenesis were based on
baryon number and $CP$ violating processes of GUT theories. The necessary
departure from thermal equilibrium was achieved by having superheavy
bosons decay by slow interactions that make them overabundant (in
comparison to their thermal distribution) in a rapidly expanding
universe \cite {weinberg, ttwz}.

It was however realized subsequently \cite {BminusL} that anomalous baryon
number violation in the electroweak theory itself \cite {thooft} 
could wipe out any baryon asymmetry formed at
the GUT scale unless the density of baryons minus leptons ($B-L$) is non
zero. Another difficulty with GUT scale baryogenesis is
inflation. Inflation is needed to get rid of heavy monopoles formed during
GUT scale symmetry breakings, but it also inflates away any baryons
produced at that scale.

Since then several other mechanisms for baryogenesis have been proposed
that produce baryons at or after the electroweak phase transition. The
most notable is the mechanism of
electroweak baryogenesis \cite {cohen}. This mechanism,
however, requires a sufficiently strong first order phase transition \cite
{cohen,kaplan}. At present the question of the order of the
electroweak phase transition remains unanswered. 
If the electroweak
phase transition were weakly first order or second order, then
alternative mechanisms of baryogenesis at or below the electroweak scale
would become very attractive.

Such models can be classified broadly by asking the two fundamental
questions:

(i) What is the source of $B$ violation? 

(ii) What is the reason for departure from thermal equilibrium? 

The question of $CP$ violation is not included among the above two
criteria. Models of baryogenesis must go beyond the standard model 
to incorporate new $CP$ violating interactions. However, we do not see any
obvious way of classifying the new $CP$ violating sectors.

The usual source of $B$ violation is one of the following two:

A) The non-perturbative $B$ violation in the standard model.

B) Perturbative $B$ violation in an extended standard model.

We would broadly classify the reason for departure from thermal
equilibrium into:\footnote {Other mechanisms
have been suggested. For example see ref. \cite {dd}.}

a) Out of equilibrium decay of massive excitations.

b) The expanding or collapsing wall. (The term ``wall''
refers to a sharp change in the
value of the Higgs field's vev.)

 The mechanism of electroweak
baryogenesis falls into the classes A) and b) respectively. The ``wall''
is the expanding wall of a bubble of true vacuum at the electroweak
phase transition. If we are looking for alternatives to electroweak
baryogenesis but want baryogenesis to occur after the electroweak phase
transition, then we either need to give up one or both of A) and b) or
look for a model that falls in the classes A) and b) but does not
require a first order phase transition.

There are in fact two mechanisms in the literature
 that are similar to electroweak baryogenesis
in that they fall in the classes A) and b) but the ``wall'' is provided
not by an expanding bubble of true vacuum but by a collapsing cosmic
string. In one of them \cite {brandenberger}, anomalous baryon number violating
processes take place inside collapsing cosmic strings. 
In another case
\cite {barriola},  electroweak strings (or $Z$ strings \cite {nambu})
with magnetic $Z$
flux are needed. 
Both these models need new $CP$ violating sectors. The first one also 
requires that sphaleron effects be appreciable inside the ``core'' of 
the cosmic strings. All models of this kind are therefore sensitive to
the structure and properties of the strings. In some cases they may
require a light Higgs which implies a first order phase transition so the
question of finding an alternative to electroweak baryogenesis remains
open. The second model needs metastable $Z$ strings, for which no viable
extension of the standard model exists so far.

The limitations of the above models lead us to explore other
alternatives to electroweak baryogenesis where we give up one or both of
A) and b). The GUT scale baryogenesis from heavy boson decays
 in fact falls
under the classes B) and a). These models automatically have the virtue
of being insensitive to the order of the electroweak phase transition. 
The question arises, whether there are viable models falling
under these classes that 
can produce baryons {\it {below}} the scale of the 
electroweak phase transition.

Seemingly, there are two hurdles to having this
mechanism work at such low energy scales. 
The first is the constraint from proton
decay. The stability of the proton implies that baryon number violating
interactions must couple to the first generation quarks with a very
small coupling. It
seems that to make this ratio small the baryon number violating
interactions must involve superheavy particles
(mass $>10^{16}$GeV). The second obstacle is that 
for a massive excitation to decay out of equilibrium, its decay (and
annihilation) rates should
be smaller than the expansion rate of the universe. With $SU(3)_C \times
SU(2)_W \times U(1)_Y$ couplings or with Yukawa couplings of the order of 
$10^{-3}-10^{-5}$, the decay (or annihilation) rates are usually greater
than the Hubble expansion rate unless the universe is at 
temperatures as high as $10^{10}$ GeV. In ref. \cite {chh} these two
hurdles were overcome by a model where proton decay was forbidden by
lepton number conservation and some heavy excitations were required to
be $SU(3)_C \times SU(2)_W \times U(1)_Y$ singlets. The model needs colored
scalars as well as a pair of massive Majorana fermions which have no
gauge interactions. 
In this paper we seek an alternative particle decay 
mechanism with, what we believe, a simpler spectrum that may still have
distinct experimental signatures.

Our motivation is in fact
twofold. Firstly we would like 
 to incorporate the merits of boson decay models
in a model that can be completely described by an effective theory at 
the electroweak scale. By bringing down the scale of $B$ and $CP$ 
violation to
the electroweak scale one improves the testability of the theory
compared to GUT scale models.
We would also 
like to restrict the fermion content to simple sequential families 
and would not
require any of them to be gauge singlets. Naturally
the model should be viable even if the electroweak phase transition is
second order. 

Secondly, extensions of the standard model, such as
models of top-color assisted technicolor \cite {hill},
suggest the possibility of having symmetries under which quarks of
different families transform differently. These symmetries must be
broken above the electroweak scale to permit quark mixing at low
energies. However the existence of these symmetries opens up the
possibility of having very small quark mixing angles. Thus the problem of
proton decay that must be addressed in models of
baryogenesis through perturbative $B$ violation
may find a new solution through small mixing angles ($10^{-4}-10^{-6}$)
between new heavy quarks that have $B$ violating interactions and the light
quarks that constitute the proton. If a viable model of baryogenesis
makes use of this mecahnism, the extra fermions needed in the $CP$
violating sector would simply 
consist of copies of the observed fermions, yet can have
distinct signatures in collider experiments which would point strongly to an
underlying theory with perturbative baryon number violation at low
energies. 

In this paper we show that the idea of small mixing angles 
described above does indeed provide a natural 
answer to the needs of any model of baryogenesis based on light boson decays.
In our model, although baryogenesis happens through the
decay of scalar bosons, the departure from thermal equilibrium can be
obtained only by having topological defects like 
cosmic strings. This is a consequence of making the theory insensitive to
the order of the electroweak phase transition \cite {ewps}.
However, we show that once the
simplest $B$ and $CP$ violating sector incorporating the above ideas is
constructed, the cosmic strings can be obtained as an inevitable bonus.
In addition, the mechanism is not sensitive to the structure and
properties of the cosmic strings.

The paper is organized as follows. In section 2 we lay out the basic
picture of how the model works. In particular we point out the various
parts of the mechanism that one must check carefully to compute the
baryon asymmetry generated in the end. The phenomenological viability of
the model is then shown part by part in the following sections.

In section 3 we describe the $CP$ and $B$ violating interactions 
and review the boson decay mechanism of
\cite {weinberg} that forms the core of the present mechanism. 
We then consider phenomenological constraints on the mixing angles of 
the model from proton decay experiments. Finally we comment on the
potentially observable experimental signatures predicted by the model.

In section 4 we present an estimate of the number density of
the scalar bosons generated by
the decay of cosmic strings in this model.

In section 5 we consider the Boltzman's equations for the evolution of
baryon number after the electro-weak phase transition and show that for
a particular range of parameters the baryon asymmetry generated
immediately after the phase transition can survive till the present
time.

\section {Basic Mechanism}

Our starting point is the boson decay mechanism. Simplest models of
baryogenesis need at least two bosons with $B$ and $CP$ violating
interactions. In the next section we will give a brief review of this
mechanism. As mentioned earlier, for successful baryogenesis there must
be departure from thermal equilibrium. However our primary motivation is
to avoid having superheavy scalar bosons. In the usual
picture of cosmological evolution, TeV scale excitations
do not go out of thermal equlibrium unless they are
practically non-interacting. Therefore we are led to consider the
following scenario. Suppose cosmic strings are formed at or slightly
above the electroweak phase transition. Some of these strings will be in
the form of loops which will subsequently decay into particles. If a
large number of these loops decay into scalar bosons that are
sufficiently heavy, then there may be an overabundance of the scalar
bosons. The decay of the overabundant scalar bosons can then be the
reason for the necessary departure from thermal equilibrium.

We would like to construct the minimal $B$ 
violating sector of scalar bosons
of the above kind. As shown in ref. \cite {weinberg}
there will be at least two of these bosons. If they
have direct $B$ violating couplings to light quarks then we have a large
width for proton decay unless the couplings are very small. But if the
couplings are very small then we do not seem
get sufficient $B$ violation and there is no baryogenesis.

A possible solution of the above problem is to have a fourth generation
of quarks. The fourth generation
can also be used to include new $CP$ violation which is a
necessary ingredient of baryogenesis.
The scalar bosons can decay into the new quarks through $B$ violating
processes leading to 
baryogenesis. However the fourth generation quarks must mix with the
        lighter generations for baryogenesis to occur in the visible
generations. In the next section we show that it is possible
to have small mixing angles ($\le 10^{-4}$) 
between the $B$ violating fourth
generation and the other three generations such that the
width for proton decay is within experimental bounds.

Thus the basic picture is the following. Immediately after the
electroweak phase transition there must be a network of decaying string
loops that produce a thermally overabundant quantity of scalar
bosons. These scalar bosons decay into a fourth family of quarks and
leptons while generating a baryon asymmetry. Finally the fermions of the 
fourth family
must decay into fermions of the other families. This is the
minimal structure for baryogenesis from light scalar decays after the
electroweak phase transition.

There are several cosmological and phenomenological constraints that the 
model must satisfy. Here we list the main constraints that
will be shown to be satisfied by the model in the next sections.

(i) All couplings must be natural. We do not address the
hierarchy problem due to scalars that also exists in the standard
model. 

(ii) Fermions of the fourth family must have experimentally viable masses.

(iii) There should not be a large width for proton decay. This implies
small mixing angles between the fourth and the other families.

(iv) Cosmic strings should be naturally incorporated in the theory.

(v) All heavy particles (the scalar bosons and the members of the fourth
family) must have large decay widths so that they decay within a
cosmologically acceptable period. A long lived heavy particle may cause
the problem of the overclosing of the universe. In particular, the
fourth generation leptons must mix with the other three generations
in order to decay. The lepton mixing must be large enough from this
point of view. 

(vi) The baryon asymmetry, once created, should not suffer a wash out
from inverse decays. 

Each of the above constraints is fairly restrictive. Indeed it is
not obvious that (iii) and (v) can be satisfied
simultaneously. However, as we show now, the minimal model built with
the motivations that we have mentioned earlier satisfies all 
the above conditions for viability. The rest of the paper is devoted to
the viability proof of this mechanism.


\section {$B$ Violation and a Fourth Generation}

\subsection {$B$ and $CP$ Violation}

The minimal light scalar boson decay model must have:

(i) two scalar bosons $X_1$ and $X_2$; 

(ii) a fourth family of quarks and leptons 
 $(\nu _4,e_4)_L, \nu_{4R}, e_{4R}, 
(t^{\prime}, b^{\prime})_L, t^{\prime}_R, b^{\prime}_R$. 

Note that
we have a right handed neutrino in the fourth family which is required
by experimental bounds on the number of light neutrino flavors ($2.99
\pm 0.04$) in the standard model. 
We can have all the necessary $B$ and $CP$ violation with the above
particle content if $X_1$ and $X_2$ are $SU(2)_W$ singlets and $SU(3)_C$
triplets and have hypercharge $Y=-2/3$.
In addition to the coupling of all the fourth generation quarks and
leptons to the standard Higgs, we can then 
have the following new Yukawa couplings:

\bear \label {lagrange}
 {L_Y} = ...g_{l1}t^{\prime}_Rb^{\prime}_RX_l + g_{l2} \overline {b^{\prime}_R}
\overline {\nu_{4R}}X_l + g_{l3} \overline {t^{\prime}_R} \overline
{e_{4R}} X_l +
g_{l4} \overline {l_{4L}} \overline{q_{4L}}X_l
+ g_{l5} q_{4L}q_{4L}X_l \nonumber \\ \vspace {2mm}
 +  {\rm {h.c}} +  ....
\eear
with $l=1,2$ and $q_4$ representing the new quark doublet.

With these interactions the action has a non anomalous $\sum
_{i=1}^3(B-L)_{i}  \times
(B-L)_4$ symmetry, where $\sum _{i=1}^3(B-L)_i$ is the difference of the baryon and
lepton numbers in the first three generations and $(B-L)_4$ is the
difference of the baryon and lepton numbers in the fourth generation.

The $CP$ violation in $L_Y$ can be communicated to the ordinary quarks and
leptons if the quarks and 
leptons of the fourth generation mix with the other generations. Note that
lepton mixing is allowed since the neutrino in the fourth generation is
massive. When these mixings are present, the quarks and leptons of the
lighter generations must couple directly to the colored scalars
$X_1$ and $X_2$ and the global non-anomalous symmetry is reduced
to $(B-L)$. We require that terms in the action that violate $(B-L)_4$
by $n$ units are suppressed by small couplings and mixings of the order of 
$s^n$ where $s$ is a small number $\sim (10^{-4}-10^{-6})$. One way to motivate this suppression
is to think of this action to be a low energy limit of a theory that has
a $(B-L)_4$ violating sector involving only massive fields. When these
massive fields are integrated out, the resulting $(B-L)_4$ violation is
suppressed by powers of their mass. In Appendix A we describe an
explicit way of realizing this suppression by considering a simple 
theory with
$gauged$ $(B-L)_{4}-\sum_{i=1}^3(B-L)_i$. When this symmetry is
broken, a low energy theory similar to the one described above is
obtained.
Other allowed couplings are gauge invariant quartic couplings
between the new scalars and the standard Higgs. The suppression rule
described above should apply to all these couplings.

The interactions  (\ref {lagrange})  violate baryon and lepton numbers while preserving
$B-L$. The interactions are in fact identical to the scalar
boson interactions of ref. \cite {weinberg}, where it was shown that the
out of
equilibrium decays of $X_1$ and $X_2$ produce a baryon asymmetry.
The amount of $CP$ violation coming from the above terms was calculated
in \cite {weinberg} and we briefly recapitulate the main results. 
At tree level, 
decays of $X_1$ and $X_2$ produce no $CP$ violation since the
cross section for $X_1 \rightarrow baryons$ is exactly equal to the
cross section for $\overline {X_1} \rightarrow antibaryons$. The same
statement applies to decays of $X_2$ and $\overline
{X_2}$. However at the one loop level there are several other
processes contributing to the same decay modes. Consider for instance
the decay $X_1 \rightarrow \overline {t^{\prime}_R}+ \overline
{b^{\prime}_R}$ and $X_1 \rightarrow t^{\prime}_R+ \nu_{4R}$.
 The one loop diagrams contributing to the process have an
internal $X_2$ propagator (Figure 1). The interference of the tree order and
one-loop diagrams produces a net $CP$ violation for the decay of
$X_1$ through these channels. The baryon number produced per decay
through the channel
`$ X_1 \rightarrow t^{\prime}_R+ \nu_{4R}$' is \cite {weinberg}:
\be \label {cp}
\Delta B_R = {4\,{\rm {Im}}(g_{11}g^*_{21}g^*_{12}g_{22}){\rm {Im}}(I) \over
  g_{11}g^*_{11}} \, ,
\ee
where $I$ is the relevant loop integral. If all the fermions have
similar masses, the loop integrals in all the channels are the same.
Taking all channels into account one then has the expression
\be \label {deltab}
\Delta B = |g|^2 {\rm {Im}}(I)\epsilon \, ,
\ee
where  $g$ is a typical Yukawa
coupling and $\epsilon$ is a phase angle characterising the average
strength of the $CP$ violation. The imaginary part of $I$ is
easily evaluated when all the fermions are massless:
\be \label {imagint}
{\rm {Im}}(I) = \left ( 16 \pi [1- \rho ^2 \rm {ln}(1+{1 \over \rho ^2})]
\right )^{-1} \, ,
\ee
where $\rho = m_{X_1}/m_{X_2}$ is the ratio of the masses of the
two colored scalars. Clearly the $CP$ violation is zero if the two scalars
have the same mass. For comparable but unequal masses \  ${\rm {Im}}(I) \sim
(10^{-2}-10^{-3})$. The value of ${\rm {Im}}(I)$ decreases if the fermions are not
massless, but the order of magnitude estimate is unchanged. 

The value of $\epsilon$ can be as large as 1 radian. Once a range of
values for the masses and mixing angles are found, the allowed range of
values for $\epsilon$ is fixed by the baryon to entropy ratio generated
in the theory.

\subsection {Masses, Mixing Angles and Proton Decay}

We shall see later that for
baryogenesis to be successful it must be possible for 
the fourth generation quarks and leptons to decay into lighter quarks
and leptons, while the reverse process must be prohibited. Therefore
all fourth generation fermions need to be heavier than the $Z$, 
so that they can
decay to a $W$ or a $Z$ and a light fermion. 
The mass difference between two members of an electroweak
doublet can not be large by considerations of the $\rho$ parameter. In
this model, since all masses come from standard couplings to the Higgs,
there is no obstruction to satisfying this criterion. Masses of the scalars
$X_1$ and $X_2$ 
need to be slightly higher than the electroweak scale. A mass of a
few TeV seems to be necessary for sufficient baryogenesis.

The mixing of the fourth generation quarks and leptons  to members of
the other three generations
provides a way for the heavy quarks and leptons to decay. As we shall see later, 
with too small mixings baryogenesis never occurs in the visible sector.
On the other hand if the mixings are too
large, the decay width of the proton increases beyond experimental
bounds. We show that there is an allowed region in the space
of the mixing angles where baryogenesis is achieved while having
acceptably small decay width for the proton.

In GUT theories where quarks of the first generation have baryon number
violating couplings, the proton decays by processes shown in Figure 2. To
meet the experimental limit on the lifetime of proton ($> 10^{33}
$years), the mass of the internally propagating GUT boson should be
$> 10^{16}$ GeV.

In the present model $B$ violating couplings involving the first three
generations are down by a suppression factor. For instance, if the
mixing angle between a fourth generation quark and a first generation
quark is $\Theta_{q_4q_1}$, a typical baryon number violating coupling
involving first generation quarks will be down by two powers of this
mixing:

\be \label {mixing}
L_Y = .... + g_{11} \Theta_{q_4q_1}^2 u_{R}d_{R}X_1 + ....
\ee

We retain the coupling $g_{11}$ to exhibit that the couplings now are
smaller in comparison to the similar couplings involving only fourth
generation quarks.
The processes leading to proton decay still look like the one
sketched in Figure 2, except the baryon number violating vertices now have very
small coupling constants. The amplitude for the process is

\be \label {amp}
{A} \sim g^2 \Theta_{q_4q_1} ^3 \Theta_{l_4l_3} {1 \over m_{X}^2} \, ,
\ee
where $\Theta_{q_4q_1}$ is a typical mixing angle 
in the quark sector  (between the first and fourth generations),
$\Theta_{l_4l_3} $ is the largest mixing angle in the lepton sector and 
 $g^2$ is the product of the typical coupling constants at
the two baryon number violating vertices {\it { without the suppression factors.}}

Using $m_{X} \sim 10^3$ GeV, we find the following inequality
for the mixing angles if experimental bounds on proton decay are to
be satisfied:

\be \label {thetalimit}
|g^2 \Theta_{q_4q_1}^3 \Theta_{l_4l_1}| \le 10^{-28} \, , 
\ee
In principle it is possible to
have greater quark mixings if the lepton mixing is smaller. Also,
$\Theta_{q_4q_3}$ and $\Theta_{q_4q_2}$ are likely to be greater than 
$\Theta_{q_4q_1}$ by one and two orders of magnitude respectively.  

The spectrum of masses and the mixing angles allowed for the
fourth family makes for interesting and distinct experimental signatures.
The presence of a sequential fourth generation is not ruled out by
experiments. The {D\O}  collaboration puts a limit of $m_{q_4} > 131$ GeV
from the charged current (CC) decay modes of the fourth generation quarks
$t^{\prime}$ and $b^{\prime}$ ($t^{\prime} \to b + W,\, b^{\prime} \to t +
W$). However if the $b^{\prime}$ is lighter than the top quark but
heavier than the $Z$, then its dominant decay mode is the flavor changing
neutral current (FCNC) mode $b^{\prime} \to b + Z$  \cite {ah}. A search for
$b^{\prime}$ through this mode is currently feasible and one expects
some experimental results in the near future (see ref.\cite {D01}). 

The $\nu_4$ is also potentially observable through the tri-lepton decay
mode $\nu_4 \to l \nu \mu$. The LEP I bound on the heavy neutrino mass
remains $m_{\nu_4} > 46$ GeV. Some restrictions on the mixing angles of
the neutrino have been placed by the {D\O} collaboration 
 \cite {D02} which are consistent
with the mixing angles allowed in the present model.
As we shall show
later, the mixing angles between the fourth and the third generation
quarks and leptons 
can be in the range $ 10^{-13} \le \Theta _{q_4q_3}\approx \Theta
_{l_4l_3} \le 10^{-4}$. Very small mixing angles will make the lightest fourth
generation quark effectively stable inside the detector and may lead to
peculiar signatures. Even if the mixing angles were 
measured to be closer to the upper bound of the above range, 
the obvious natural explanation for
their smallness would be the existence of extra symmetries at scales
higher than $1$ TeV which forbid the mixing between the fourth
generation and the other generations. Then, with the standard gauge
and Higgs couplings, there is an extra symmetry in the form of $B_4 -
L_4$ above the TeV scale. For the small quark mixings in the fourth
generation to exist, this symmetry
must be broken at the TeV scale and a likely result would be
perturbative violation of
baryon number in the fourth
generation. Therefore experimental
signatures that are consistent with the mixing angles predicted 
in this model point very strongly to a
mechanism of baryogenesis through perturbative $B$ violation.

\section {Cosmic Strings and Baryogenesis}

With the $B$ violation and $CP$ violation in place, all we need to produce
baryons is an out-of-equilibrium decay of the scalar bosons. To achieve
this there has to be a mechanism for making the scalar bosons
overabundant immediately after the electroweak scale. 
This can happen through the formation and decay of cosmic strings. The
cosmic strings must form close to the electroweak phase transition. Strings
formed much earlier will have a distribution with a correlation length
that is too large and 
their number density will be too small to generate the baryon excess
we see today. Since the symmetry
breaking in the standard model does not produce any cosmic
strings, extra broken gauge symmetries must exist.
However, as described in Appendix A, the broken symmetry can be the
approximately conserved $(B-L)_4 - \sum _{i=1}^3(B-L)_i$. If this
is a gauged symmetry that is spontaneously broken at the electroweak
scale, the
smallness of the mixing angles and the presence of the cosmic strings
can be explained simultaneously.

Moreover, besides being candidates for seeding galaxies, cosmic strings 
occur in many, independently motivated extensions of 
the standard model. 
For instance the Aspon
model \cite {aspon} has an extra $U(1)$ that breaks close to the electroweak
scale. The motivation there is to provide an explanation for the small
value of the vacuum angle $\theta$ in QCD. Supersymmetric extensions of
the standard model have been proposed which attempt to resolve the $\mu$
problem of the minimal supersymmetric standard model (MSSM) and the
cosmological solar neutrino problem and which have strings \cite {mssm}. Both
these models were considered recently \cite {mark} in the context of
electroweak baryogenesis from cosmic strings. Top color models \cite {hill} are
another class of models where a $U(1)$ gauge group is broken close to
the electroweak scale. In our case, almost
regardless of motivation, any extension
of the standard model that produces strings at the electroweak scale
will work. The cosmic strings need not have a special structure or 
satisfy any particular
requirements. The effective scalar whose vacuum expectation value 
(vev) causes the formation of the
cosmic strings will naturally have quartic couplings with the bosons
$X_1$ and $X_2$. This ensures that $X_1$ and $X_2$ will be produced from
the decay of cosmic string loops.

The scenario for baryogenesis is now very similar to \cite {branden} where the
authors considered emissions of heavy particles from collapsing cosmic
strings. If the heavy particles are produced at a scale which is
sufficiently small compared to their mass, they may become overabundant and through
$CP$ violating decays generate baryon number. In the present model we
focus on the overabundant production of $X_1$ and $X_2$ particles
from strings.

Immediately after
the electroweak phase transition, the space will be filled with a criss-crossing of
string network that looks like a random walk
in three dimensions. The initial correlation length is  $\psi_{t_c} \sim {1
\over T_{c}}$. Numerical simulations indicate \cite {vachaspati} that a
large fraction ($\sim 80 \%$) of the total string length resides in the
infinite strings. The rest is in the form of loops which have a scale
invariant distribution
\be \label {init}
{dn \over dR} = R^{-4} \, , 
\ee
where $R$ is the characteristic size of the loops. 
The initial network has loops which decay rapidly. The infinite segments
also generate more loops by frequent intercommutations. The net result
is that the correlation length increases with time and the string
network 
enters a scaling solution when  the correlation length equals the
horizon size \cite {branden}.

In the literature the period from the time of the phase transition
($t_c$) to the time $t^* = (G \mu )^{-1}t_c$ is called the friction
dominated period \cite {friction}. ($G$ is the Newton's constant and
$\mu$ is the mass per unit length of the strings).
Loops produced at time $t$ with $t_c < t < t^*$
immediately shrink to the radius
\be \label {rad}
r_f(t) = G \mu m_{pl}^{1/2}t^{3/2} \, , 
\ee

where $m_{pl}$ is the Planck mass ($\sim 10^{19}$GeV). 
Below this scale friction effects are subdominant and the loops shrink
predominantly by processes like gravitational radiation and cusp
annihilation. The shrinkage rate from gravitational radiation is \cite {gravi}
\be \label {grav}
{dR \over dt} = - \gamma G \mu \, , 
\ee
where $\gamma$ is numerically determined to be about 10.

The time for a loop to shrink to a size which is of the order of its
thickness due to purely gravitational effects is $t_G = (\gamma G \mu
)^{-1} R$. In the present case $t_G \sim 10^{31}R \sim 10^{18}s$ even
for the smallest loops ($R \sim \psi$). This is a very long time
compared to the cosmological times of interest and gravitational effects
are therefore completely negligible in our model.

Cusp annihilations, on the other hand, can occur at a much faster
rate. The rate of shrinkage in this case can be modelled by \cite
{cuspi}
\be \label {cusp}
{dR \over dt} = - {\gamma_{\rho} \over (R \eta)^{1/3}} \, .
\ee
The corresponding decay time is $t_{\rm {cusp}} = {R(R \eta)^b \over
\gamma_{\rho}}$. As shown in \cite {branden} this is less than one expansion time if
the time of formation $t < ($$ \gamma_\rho^2 \over G \mu $$)^{1/2} t_c$. For $G \mu \sim
10^{-32}$ it is safe to assume that this condition is satisfied for a
long time. Thus in our model, the loops formed at time $t$ immediately
shrink to $R = r_f(t)$ after which they shrink by cusp annihilation to $R
\sim \mu ^{-1/2}$ within one expansion time. The lifetime of a loop formed at
time $t$ is then $\tau(t) \sim t$.

When the string loops have shrunk to a radius that is comparable to
their thickness $\mu^{-1/2}$ (where $\mu$
is the mass per unit length of the string), nonlinearities in the scalar
field potential will cause the entire loop to decay into elementary particles.
It is from this final burst process that we can expect the heavy scalars
$X_1$ and $X_2$ to be produced. Since $\mu \sim
m_{X_1}^2, m_{X_2}^2$; the number $N$ of $X_1$ and $X_2$ particles
that we get from each loop is $ \sim 1$.

The baryon number produced by string decays can now be evaluated by
computing the number of string decays from the time $t_c$ of the phase
transition up to the present time. There are two kinds of loops to be
taken care of; those formed {$\it {at}$} the time of phase transition ($t =t_c$) and
those formed {$\it {after}$}  the phase transition ($t> t_c$). Also, the baryon
number produced from the decay of $X_1$ and $X_2$ bosons will be
washed out unless baryon number violating processes are effectively
frozen after the $X_1$ and $X_2$ decays. To compute the remnant
baryon asymmetry we therefore need to evaluate the number of string
decays that take place $\it {after}$ $t=t_f$, where $t_f$ is the freeze
out time for baryon number violating interactions. Two cases, thus,
arise:

1) $ { t_f \approx t_c }$

The number of decaying loops is the sum of loops produced at $t_c$ and
$\it {after }$ $t_c$. The number density of heavy bosons produced from
loops formed $\it {at}$ $t_c$ is obtained by integrating (\ref {init}):
\be \label {enone}
n_{t_c} \sim {N\over \psi ^3_{t_c}} \left ({a_{t_c} \over a_{t}}\right )^3 \, , 
\ee
where the last factor takes care of the dilution of the number density
due to the expansion of the universe.
More cosmic string loops are formed after the
phase transition as the coherence
length increases with time and loops are chopped off from `infinite' sections
of the string network. The loop production rate is related to the rate
at which the coherence length increases by \cite {branden}
\be \label {loop}
{dn \over dt} = {\nu \over \psi^4} {d\psi \over dt} \, , 
\ee
where $\nu $ is a constant of the order of unity. Integrating (\ref
{loop}) we have 
\be \label {entwo}
n_{t>t_c} = N \int_{\psi _{t_c}}^{\psi _{t}}{\nu \over {\psi} ^{\prime
    4}}  \left ({a_{t ^{\prime}}\over a_{t}}\right )^3 d{\psi ^ {\prime}}
 \sim N {\nu a_{t_c}^3 \over {(\psi _{t_c}a_t)}^3} \approx n_{t_c}   \, .
\ee
In order to have a large $n_{t_c}$ we need to minimize $\psi _{t_c}$.
Using $\psi _{t_c} \sim {1 \over T_c}$  for strings produced at the
electroweak scale, we get the ratio of the 
total number
density of $X_1$ and $X_2$ to the entropy density to be
\be \label {nmax}
\omega_{\rm {max}} \sim {n_{t_c} \over s} \approx {N \over g_*} \, , 
\ee
where $s$ is the entropy density of the universe and 
$g_* \sim 100$ is the number of effectively massless degrees of
freedom at the electroweak scale.

2)  $ {t_f > t_c}$

Now the loops produced at $t=t_c$ decay within one expansion time and the
resulting baryon number is washed out. Thus we need consider only the
contribution from the loops decaying after $t_f$:
\be \label {enthree}
n= N \int _{\psi _{t_f}}^{\psi _{t}} {\nu \over {\psi}^{\prime 4}}
\left ({a_{t^{\prime}} \over a_{t}}\right )^3
d{\psi ^{\prime}} \sim n_{t_c} {\psi _{t_c}^3 a_{t_f}^3 \over \psi
  _{t_f}^3 a_{t_c}^3} \, .
\ee
In the friction dominated period $\psi \sim t^{5 \over 2}$ \cite
{branden}, therefore we have, 
$\omega = \omega_{\rm {max}} ({t_c \over t_f})^3$. Thus unless $t_f = t_c$,
there is some damping in the production of baryons. In the next
section we show that it is possible to have $t_f \approx t_c$ in our
model. 

\section {Approach to Equilibrium}

The baryon asymmetry generated from $X_1$ and $X_2$ decays will
evolve according to the Boltzman's equations. A large number of
particles and interactions are relevant. For each species of particles
one gets a Boltzman's equation. The
equations are coupled integro-differential equations and for an accurate estimate
of the baryon number, one must integrate them numerically.
Useful analytical estimates can, however, be made by making simplifying
approximations that reduce the number of degrees of freedom (and hence
the number of Boltzman's equations) to a few.

The greatest simplification results from considering a single
scalar boson $X$  instead of $X_1$ and $X_2$. In the following
we consider the most dominant interactions of this scalar boson. 
The relevant processes are:

{\halign{\hfil\bf#&\quad#\hfill\cr

1.&$X \overline{X} \rightarrow q \overline{q},GG,l\overline {l} \, {\rm {etc}}.$\cr
2.&$X \rightarrow q_4 q_4$\cr
3.&$X \rightarrow \overline {q_4}\overline{l_4}$\cr
4.&${q_4}{q_4} \rightarrow \overline {q_4} \overline
{l_4}$\cr
5.& $q_4 \rightarrow q_3 W$\cr
6.& $l_4 \rightarrow l_3 W$\cr
}

$X $ is the generic colored scalar; $q$ and $l$ refer to generic
quarks of any generation while 
$ q_4$ and $l_4$ refer to a
quark and a lepton of the fourth generation. The $W$ in
interactions 5 and 6 represents the $W$ bosons of the weak interactions. 
We denote gluons by $G$.

The processes 1 are dominant annihilation channels for the colored
scalars. 
The processes 2,3,4 are baryon number violating processes. The
processes 2,3 generate the baryon number while their inverse processes
and processes 4 can                       
wash out the produced baryon number. 
The processes 5 and 6 transport baryon and
lepton number from the fourth generation to the third generation. 
We have omitted processes of the kind $q_4 \rightarrow \overline {q_4}
\overline {l_4}\overline {q_4} $ to simplify the Boltzman's equations. These
processes are prohibited if the quarks and leptons of the fourth generation
have nearly equal masses. Inclusion of these processes does not change
our main results in a significant way. 
The processes of baryogenesis and `freeze out' can now be treated as two
distinct stages in the evolution of the baryon number.

\subsection {$X$ production and decay}

We will assume that immediately after the phase transition the decays of
cosmic string loops raise $n_X$,the number density of $X$, to about
$T_c^3$. (Later we show that $n_{X} > 10^{-2} T_c^3$ may be sufficient
for baryogenesis).
Since $M_X > T_c$, the $X$s are 
overabundant and their number will decrease rapidly through decays and
annihilations. 

The Boltzman's equation for $X$ is \cite {kolbturner}

\bear \label {boltchi}
{dn_{X} \over dt}+3Hn_{X}= -\int DP_{X,ij}\left [f_X |M(X \rightarrow ij)|^2 -
f_if_j|M|(ij \rightarrow X)|^2 \right ] \nonumber \\ \vspace{2mm}
 - \int DP_{X \overline {X},ij}\left [f_Xf_{\overline {X}} |M({X \overline
  {X}} \rightarrow ij)|^2 -
f_if_j|M(ij \rightarrow {X \overline {X}})|^2\right ] \, , 
\eear
where $DP_{a_1a_2...,b_1,b_2..}=
\Pi _{i}\int {d^3 {\bf p}_{a_i} \over (2 \pi)^3 2E_{a_i}}
\Pi _{j}\int {d^3 {\bf p}_{b_j} \over (2 \pi)^3 2E_{b_j}}
(2\pi )^4\delta ^4(\Sigma _ip_{a_i}-\Sigma _jp_{b_j}) $ is the phase
  space volume
element, $H$ is the Hubble constant,
$f_i$ is the phase space density of species $i$ and $|M(ij
.. \rightarrow ..kl)|^2$ is the matrix element squared for the process
$ij.. \rightarrow ..kl$. The matrix element is summed over initial
and final state color, spin and flavor degeneracies. The number density
$n_X$ is the number density of all $X$ particles regardless of color.

When the final state particles are light, (\ref {boltchi})  reduces to
\cite {kolb} \, :

\be \label {newchi}
{dn_{X} \over dt}+3Hn_{X}= -\left [n_{X}-n^{\rm {\rm {eq}}}_X\right ]\langle \Gamma_X \rangle
-\left [n_{X}^2-({n^{\rm {eq}}_X})^2 \right ]\langle \sigma (X \overline X \rightarrow ij)\rangle
\ee
where $\langle\Gamma_X\rangle$ is the total thermally averaged decay width of $X$
averaged over initial color degeneracies,\ $n^{\rm {\rm {eq}}}_X$ is the
equilibrium density of $X$ and $\langle \sigma (X \overline X \rightarrow
ij)\rangle$
is the thermally averaged cross section for $X \overline X$
annihilations (averaged over initial state color degenaracies).

The decay modes for $X$ are: $X \rightarrow q_4q_4, \overline {q_4}
\overline {l_4}$. The dominant annihilation channels are: $X\overline X
\rightarrow q \overline q, GG$. Since $X$
is very heavy we can use the zero temperature decay width for $\Gamma_X$ to good
approximation. The same is true for the annihilation cross section with
an appropriate value for the c.m energy. In Appendix B we have computed
these rates. Our results are:
\bear \label {decaywidth}
\langle \Gamma_X\rangle \approx  {1 \over 2\pi} |g|^2M_X  \, , \\ \vspace{2mm}
\langle \sigma (X \overline X \rightarrow ij)\rangle \approx {(4 \pi 
\alpha_{\rm {qcd}})^2 \over 9 \pi M_X^2} \, , 
\eear
where $g$ is a typical coupling of $X$ to $q_4$ and $l_4$.

Because $X$ is overabundant, $n_X \gg n^{\rm {\rm {eq}}}_X$. The reduction in $n_X$
is initially dominated by annihilation processes. The decays overtake
annihilations when the $X$ number density reaches the critical value
\be \label {ncrit}
n_{X\rm {crit}} = {9|g|^2 \over 2} {M_X^3 \over (4 \pi \alpha _{\rm
    {qcd}})^2} \, .
\ee
{}From this point onwards, the annihilations are quenched out and most of
the $X$s decay through the $CP$ violating processes producing baryons.

\subsection {Freeze out}

Once a large number of $X$ decays have taken place, there is an excess
of baryons (and leptons) over anti-baryons (anti-leptons) in the fourth
generation. Baryon number violating processes like inverse decays
($q_4q_4, \overline {q_4} \overline {l_4} \rightarrow X$) or $2
\rightarrow 2$ processes ($q_4q_4 \rightarrow  \overline {q_4} \overline {l_4}$) will tend
to wash-out this excess. Decays to $W$s and third generation quarks and
leptons will also reduce the baryon excess in the fourth generation
(although preserving the total baryon excess). 

To see if a freeze out can occur, we look at the Boltzman's equation for
$q_4$:
\bear 
{dn_{q_4} \over dt} + 3Hn_{q_4} &=&-3 \int DP_{q_4q_4,\overline {q_4}\overline
{l_4}} \left [f_{q_4}f_{q_4} |M^{\prime}(qq \to \overline {q_4} \overline
{l_4})|^2 - f_{\overline {q_4}}f_{\overline {l_4}}|M^{\prime}(\overline
{q_4} \overline{l_4} \to qq)|^2 \right ] \nonumber 
\\ \vspace{2mm}
&+&2\int DP_{X,q_4q_4} \left[f_X |M(X \to q_4q_4)|^2 - f_{q_4}f_{q_4}|M(q_4q_4 \to
X)|^2 \right] \nonumber
\\ \vspace{2mm}
&+&\int DP_{\overline {X},{q_4}{l_4}} \left[f_{\overline {X}}|M(\overline {X} \to {q_4}
{l_4})|^2 - f_{q_4}f_{L_4}|M({q_4}{l_4} \to \overline {X})|^2\right]  \nonumber 
\\ \vspace{2mm}
&-& \int DP_{q_4,q_3W} \left[f_{q_4}|M(q_4 \to q_3W)|^2 -
f_{q_3}f_{W}|M(q_3W \to q_4)|^2\right] \, .
\label {boltq}
\eear

$|M^\prime(q_4q_4 \rightarrow \overline{q_4}\overline{l_4})|^2$ and 
$|M^\prime(\overline{q_4}\overline{l_4}
\rightarrow q_4q_4)|^2$ have primes on them to indicate that the matrix
elements do not include {\bf s} channel contributions in which the
intermediate $X$ is on shell (a physical particle), since these
contributions have already been included in the decay and the reverse
decay terms. The full matrix element (squared), $|M(q_4q_4
\rightarrow q_4l_4)|^2$, is related to $|M^\prime(q_4q_4 \rightarrow q_4l_4)|^2$
by
\bear \label {matsq}
&&|M(q_4q_4 \rightarrow \overline{q_4}\overline{l_4})|^2 = |M^\prime(q_4q_4
\rightarrow \overline{q_4}\overline{l_4})|^2 \nonumber
\\ \vspace{2mm}
&+&{\pi \over M_X \Gamma_X} \delta \left[p^2_{q_4}(1) + p_{q_4}^2(2)
-m_X^2\right]
|M(q_4q_4 \to X)|^2 |M(X \to \overline {q_4}
\overline {l_4})|^2 \, .
\eear

Following \cite {kolb}  we will simplify (\ref {boltq})  by parametrizing the $CP$
violation of the system in the following manner. We define the matrix
elements $M_0$ and the numbers $\eta$ and $\overline \eta$ by
\bear \label {etaetabar}
|M(X \to q_4q_4)|^2 =& |M_0|^2(1+\eta)/2 &= |M(\overline {q_4}^{\,} \overline {q_4}
\to \overline {X})|^2 \, , \nonumber 
\\ \vspace{2mm}
|M(X \to \overline {q_4}\overline {l_4})|^2 =&|M_0|^2(1-\eta)/2 &= |M({q_4}{l_4}
\to \overline {X})|^2 \, , \nonumber
\\ \vspace{2mm}
|M(\overline {X}\to \overline {q_4}^{\,} \overline {q_4})|^2 =& |M_0|^2(1+\overline
{\eta})/2 &= |M({q_4}{q_4} \to {X})|^2 \, , \nonumber
\\ \vspace{2mm}
|M(\overline {X} \to {q_4}{l_4})|^2 =& |M_0|^2(1-\overline {\eta})/2 &=
|M(\overline {q_4} \overline {l_4}\to {X})|^2 \, .
\eear

We have used $CPT$ and unitarity to relate the squared matrix
elements. Note that all matrix elements are summed over initial and final
state spin and  color degeneracies. The $CP$ violation parameters $\eta$
and $\overline \eta$ are related to $\Delta B$ by the relation
\be \label {cprel}
\Delta B = (\eta - \overline \eta)/4 \, .
\ee

Since all the quarks and leptons are in thermal equlibrium we have
\bear 
f_{q_4}(p) = e^{-E/T +\mu_1/T} \approx e^{-E/T}\left (1+{b \over
  2}\right ) \, ,\nonumber
\\ 
f_{\overline {q_4}}(p) = e^{-E/T -\mu_1/T} \approx e^{-E/T}\left (1-{b \over
  2}\right ) \, , \nonumber
\\ 
f_{L_4}(p) = e^{-E/T +\mu_2/T} \approx e^{-E/T}\left (1+{l \over 2}\right )
\, , \nonumber
\\
f_{\overline {L_4}}(p) = e^{-E/T -\mu_2/T} \approx e^{-E/T}\left (1-{l
  \over  2}\right ) \, , 
\label {phasespace}
\eear
where $\mu_1$ and $\mu_2$ are chemical potentials related to the
(approximately) conserved baryon and lepton numbers in the fourth
generation. In expanding the exponents we have used the fact that baryon
and lepton excesses are small. {}From (\ref {phasespace}) we obtain
\bear \label {enbee}
\sum _{g_i=1}^2 \sum _{s=1}^2 \int {d^3{\bf {p}} \over (2
\pi)^3} f_{q_4}(p)f_{\overline {q_4}}(p)= {B_4} ~, \nonumber 
\\ \vspace{2mm}
\sum _{g_i=1}^2 \sum _{s=1}^2 \int {d^3{\bf {p}} \over (2
\pi)^3} f_{l_4}(p)f_{\overline {l_4}}(p)= {L_4} ~,
\eear
which relate $b$ and $l$ to the density of excess baryons and leptons
respectively. The sums are over the flavor and spin indices in the fourth
generation. 

One can now use  (\ref {matsq})  and (\ref {phasespace})  in (\ref
{boltq})  and express products
like $f_{q_4}(p_1)f_{q_4}(p_2)$ as $f^{\rm {eq}}_{X}(p_1+p_2)(1+2b)$ in decay
and inverse decay terms. Subtracting the Boltzman's equation for
the antiquarks from the equation for the quarks one gets the equation
for the baryon number in the fourth generation:
\bear 
{d{B_4} \over dt} + 3H{B_4} & \approx & {1 \over 2}(n_X-n_X^{\rm {eq}})(\eta -
\overline {\eta})\langle \Gamma(X \to q_4q_4,\overline {q_4} \overline
{l_4})\rangle
\nonumber \\
&-& {1 \over 4}(3{B_4}+ {L_4})n_{\gamma}\langle \sigma(q_4q_4 \to \overline {q_4}
\overline {l_4})\rangle 
\nonumber \\
&-& {1 \over 6}{n_X^{\rm {eq}}\over n_{\gamma}}\left [{1 \over 2}(B_4+L_4)\langle \Gamma(X \to q_4l_4)\rangle + B_4  \langle \Gamma(X \to
q_4q_4)\rangle \right ]
\nonumber \\
&-& {1 \over 2} (B_4-B_3) \langle \Gamma(q_4 \to q_3 W)\rangle \, , 
\label {baryonfour}
\eear
where $\langle\Gamma(X \rightarrow \overline {q_4} \overline {l_4})\rangle$ 
and $\langle\Gamma(X \rightarrow q_4q_4)\rangle$ are averaged over initial state degeneracies and summed over
final state degeneracies, $\langle \sigma (q_4q_4 \to \overline {q_4}
\overline {l_4})\rangle $ is summed over $both$ initial and final state
degeneracies and $n_{\gamma} \approx T_c^3$ is the photon number density.
The correct sign for
the term $-n_{X}^{\rm {eq}}(\eta - \overline {\eta })\langle\Gamma(X \to
q_4q_4,\overline {q_4} \overline {l_4})\rangle$ is obtained
only after including the $CP$ violating part of $|M^{\prime}(q_4q_4
\rightarrow \overline {q_4} \overline {l_4})|^2 - 
|M^{\prime}(q_4q_l\rightarrow q_4q_4)|^2$
\cite {kolb}. 

The various terms in the r.h.s of (\ref {baryonfour}) are
readily interpreted. The
first term is the driving term for baryogenesis. It becomes small as
$n_X \rightarrow n^{\rm {eq}}_X$ and plays no role in freeze out.
The second term comes
from inverse decays and the third term comes from $2 \rightarrow 2$
baryon number violating processes. These two terms can potentially cause
a washout.
The last term is the rate at which
baryon number is drained out of the fourth generation into the third
generation. We have ignored similar drainage terms to other generations
because the mixing angles are smaller by one or two orders of magnitude.

The `washout terms' can be ineffective only if they are smaller than the
Hubble dilution term $3H{B_4}$. We must, therefore, have (using $
{L_4} \approx {B_4}$) 
\bear
3H{B_4}> {B_4}n_{\gamma}\langle
\sigma(q_4q_4 \to \overline {q_4} \overline {l_4})\rangle ~,
\nonumber \\ \vspace{2mm}
3H{B_4}> {1 \over 6}{n_X^{\rm {eq}}\over n_{\gamma}}B_4
\langle \Gamma(X \to q_4 q_4, \overline {q_4} \overline {l_4})\rangle \, .
\label {freez}
\eear

\subsection {Range of parameters}

At the weak phase transition $H \sim 3 \times 10^{-16}T_c$. Using our estimates
of the decay widths and cross sections from Appendix B we can reduce the
conditions (\ref {freez}) to
\bear
10^{-16} &>& {4|g|^4 \over 3 \pi k^4 } ~,
\nonumber \\ \vspace{2mm}
10^{-16} &>& {|g|^2 \over 36 \pi}k^{5/2}e^{-k} ~,
\label {freeznum}
\eear
where $k={m_X \over T_c}$.

When these inequalities are satisfied, there is no significant washout
of the baryon number and the 
net baryon to entropy ratio is
\be \label {pararatio}
\eta_B = {n_{X \rm {crit}}\Delta B \over g_* T_c^3} \sim 10^{-4}|g|^4k^3
\epsilon \, .
\ee
We have taken $(4\pi \alpha_{\rm {qcd}})^2=2$ and $g_* = 100$. 
Two parameters
in (\ref {pararatio}) are bounded from above. The maximum value of $\epsilon
\sim 1$ and the maximum value of $n_{X \rm{crit}} \sim T_c^3$. When these
bounds are taken into consideration, the inequalities (\ref {freeznum})  and
(\ref {introbar}) yield
the following range of values for $k$ and $|g|^2$:
\bear \label {kgbound}
25 &<& k \, <\,  80 \,  , \nonumber \\ \vspace{2mm}
10^{-6} &<& |g|^2\,  <\,  10^{-3.5} \, .
\eear
Picking some value for $k$ further constrains the range for $|g|^2$ and
vice versa. Realistic values for $\eta_B$ can be obtained with these
values. For instance, taking $k=30$, $|g|^2 = 10^{-5}$ we obtain,
\be \label {baryoresult}
\eta_B \sim 2.7 \times 10^{-10}\epsilon \, .
\ee
For $\epsilon$ close to 1 this falls within the range given by (\ref {introbar}).
Note that for $k \sim 25$, $m_X \sim 6.25$ TeV. This value is to be compared
with the mass of the smallest string loops. Indeed if the mass per unit
length of the strings is $\mu$, a string loop of size $R \sim {1 \over
  T_c}$ has a mass of about $\beta R \mu $, where $\beta$
is a numerical factor that takes into account the fact that loops are
not exactly circular. Numerical simulations indicate that $\beta \sim
9$  \cite {shellard}. If $\mu \sim $(TeV)$^2$ then the mass of the smallest loops is about
$36$ TeV. Also note that $\eta _B$ is insensitive to the number $N$ of
$X$s produced per string loop as long as $NT_c^3 > n_{X\rm {crit}}$. For $|g|^2
= 10^{-5}, k= 30$ we have $n_{X\rm {crit}} \sim 10^{-1}T_c^3$. 

The range of allowed values for the mixing angles is much wider. {}From
(\ref {thetalimit})  and (\ref {kgbound})  we can see that
\be \label {moremix}
|\Theta_{q_4q_1}^3 \Theta_{l_4l_3}| \le 10^{-22} \, .
\ee
Now consider the decay of the fourth generation baryons and leptons. The
decay widths are $\sim {|g_W\Theta _{l_4l_3}|^2 T_c \over 4 \pi}$ (where
$g_W$ is the weak gauge coupling). A
lower bound on the mixing angles is obtained by requiring that the
decays happen before nucleosynthesis. This means that the decay time
should at most be $1s$. The corresponding limit on 
the mixing angles is: $\Theta_{l_4l_3}, \Theta_{q_4q_3} \ge 10^{-13}$. 
If we also require that $\Theta _{l_4l_3} \approx \Theta _{q_4q_3}
\approx 10  \Theta _{q_4q_2} \approx 100  \Theta _{q_4q_1}$, then we have
\be \label {limitheta}
10^{-13} \le \Theta _{q_4q_3} \le 10^{-4} \, .
\ee

\section {Conclusions}

We have shown that baryogenesis from the decay of light scalar bosons is
viable even at energies as low as the electroweak scale. This is
interesting, since perturbative violation of baryon number at low
energies seems incompatible with the observed stability of the
proton. However the minimal model of $B$, $C$ and $CP$ violation
involving light scalar bosons can naturally have very small mixings
between new heavy quarks with $B$ violating interactions and the lighter
quarks which shields the proton from $B$ violating effects. Other 
phenomenological and cosmological constraints are shown to be satisfied.
In particular the small mixings between the fourth family and the other
families is shown to be sufficient for quarks in the fourth generation
to decay into quarks of the lighter generations in a cosmologically
acceptable time. 

Some members of the fourth family can be as light as 100 GeV. They
can also be relatively long lived (decay time $\sim 10^{-5}s$). It would
be interesting to explore signatures of their existence in future
experiments. In particular, if the lighter quark in the fourth family,
the $b^{\prime}$, happens to be lighter than the top quark then its
dominant decay mode is the FCNC mode $b^{\prime} \to b + Z$. One expects
this decay mode to be explored in collider experiments of the near
future. Signatures associated with new quarks 
with small 
mixing angles ($ \Theta _{q_4q_3} \le 10^{-4}$) as predicted by this model,
would seem to point strongly toward a mechanism of
baryogenesis through perturbative $B$ violation at the electroweak phase
transition.

The scalar bosons must be at least 25 times heavier than the electroweak
scale. In our model they are produced copiously from the decay of loops
of cosmic strings immediately after the electroweak phase transition. 
We show in Appendix A, it is possible to extend the
standard model so that the smallness of the new mixing angles and the
presence of the cosmic strings are justified simultaneously. 

Variations of this model can be conceived. The only necessary
ingredients are topological defects like cosmic strings and heavy
baryons. The model has all the advantages of baryogenesis models where
baryogenesis occurs $after$ the electroweak phase transition including
compatibility with the usual models of inflation.
It is also
viable as a baryogenesis model even if the electroweak phase transition
is a second order phase transition.

\vfill \eject
{\centerline {\bf {Acknowledgements}}}
\vskip 0.5cm
I would like to thank B.Balaji, J. Butler, 
R.Brandenberger, S.Chivukula, A.Cohen,
B.Dobrescu, K.Lane, T.Vachaspati and V. Zutshi 
for discussions and valuable suggestions.
This work was supported by the Department of Energy under grant
DE-FG02-91ER40676.

\vskip 2.0cm

{\centerline {\bf {Appendix A}}}
\vskip 0.5cm
The model described in section 2 has an anomaly free global
$(B-L)_4$ symmetry in addition to the ususal $B-L$ when the mixing angles between
the fourth and lighter generation are put equal to zero and there is no
coupling between the lighter generation quarks and leptons and the
colored scalars $X_1$ and $X_2$. When the mixings and couplings
mentioned above are non-zero but small, $(B-L)_4$ is broken weakly. The
smallness of these parameters is, therefore, not unnatural in the
technical sense. Below, we describe a way of explicitly realizing this
scenario as an effective low energy limit of a theory where the small
$(B-L)_4$ breaking terms come from operators of dimension 5
or higher and are suppressed by a large mass scale.

We first define two $U(1)$ symmetries:
\be \label {uvsym}
U=\sum _{i=1}^4 (B-L)_i = B-L \; ; \; \; \; \;
V= (B-L)_4 - \sum _{i=1}^3 (B-L)_i \,.
\ee
The first is the usual $B-L$ which we keep as a global unbroken
symmetry of our theory. At a scale much higher than the electroweak
scale one can conceive of a theory where $V$ is a gauged
symmetry. Consider for instance extending the model in section 3 by
first taking away all terms violating $V$ and then gauging
$V$. (The terms involving the scalar $X_3$ are not necessary for
this discussion and can be discarded).
In order that $V$ be realized as a weakly broken symmetry in the
low energy theory we can introduce a scalar field $X_3$ which is a
singlet under all gauge symmetries except $V$. A vev for $X_3$ breaks
$V$. 

The effective action for this theory can have a dimension $4+n$ operator of
the kind $ {1 \over M^n}\overline {q_4}hq_3X_3^{n}$ if the $V$ charge of
$X_3$ is
${2 \over 3n}$. $M$ is a large mass suppressing this operator ($h$ is the standard
Higgs). When $X_3$ gets a
vev, $V$ is broken and we get the dimension 4 mixing term $({<X_3> \over
M})^n \overline {q_4}hq_3$. The smallness of the mixing results from
suppression due to small coupling constants and factors of $1 \over
4\pi$ as well as the ratio $<X_3> \over M$. 

To actually get this operator we must introduce new fields and
interactions that couple the fourth and the third generations. Since we are
not interested in solving the hierarchy problem, scalars are cheap. We
can introduce three more, $X_4$, $X_5$ and $X_6$  with the following Yukawa couplings
\be \label {xfive}
L_{\rm {Yukawa}} = g^{\prime} \left (q_4q_3X_4 + \overline 
{q_4}\overline {l_3}X_5 +
\overline {q_3}\overline {l_4}X_6 \right ) \, ,
\ee
where $q_i$ and $l_i$ denote quarks and leptons of the $i$th
generation. We have chosen the same coupling $g^{\prime}$ for all the
terms. The $V$ charges of $X_4$, $X_5$ and $X_6$ are $0$ and ${4 \over
  3}$ and $-{4 \over 3}$ respectively. All
are color triplets and $SU(2)_W$ singlets. By suppressing the flavor and helicity indices we imply that all
possible gauge invariant couplings are included in (\ref {xfive}). 

Now
suppose $X_4$, $X_5$ and $X_6$ have masses of the order of $M$. Integrating
them out one may obtain the dimension 6 operator $q_3q_3q_3l_3$.
This operator, a potentially dangerous candidate for proton decay, 
is not induced by renormalization at one loop and must be 
suppressed by at least a factor of ${g^2g^{\prime 4} \over 16 \pi ^3
M^2}$ where $g$ is one of the couplings in (\ref {lagrange}). If we choose
$|g|^{\prime 2} < |g|^2 \approx 10^{-5}$, the proton decay problem is
avoided for $M^2 > 10^9$ GeV$^2$.

The mixing between the fourth and third generations can occur through the
operators $\overline {q_4}Dq_3X_3^2$ (where $D$ is the gauge-covariant
derivative) or 
$\overline {q_4}hq_3X_3^2$. Figures 3a and 3b show typical leading order
contributions to these operators. Clearly the $V$ charge of $X_3$ is ${1
\over 3}$. The mixing is suppressed by the small number ${|g||g^{\prime
}|\over (4\pi )^2} ({\langle X_3\rangle \over M})^2$. For $\langle
X_3\rangle  \approx 10^3$ GeV, $M
\approx 10^{4.5}$ GeV and
$|g||g|^{\prime}
\approx 10^{-5}$, the mixing is
$\Theta_{q_4q_3} \approx 10^{-10}$. A similar mixing is obtained in the
lepton sector. Mixings of this order are certainly small enough for the viability
of our model. The mixings are also large enough for the baryon number in
the fourth generation to be transported to the lighter generations in a
cosmologically acceptable time. 
Indeed with this mixing the decay time of the fourth generation 
quarks and leptons is $\sim 10^{-5}s$ which corresponds to a temperature
of about 1 GeV. 
By choosing $\langle X_3\rangle \sim 10^{3}$GeV we also get the much needed cosmic
strings at the electroweak scale as a bonus.

\vskip 2.0cm
{\centerline {\bf {Appendix B}}}
\vskip 0.5cm

The dominant annihilation modes of $X$ are $X \overline {X} \rightarrow
GG, q\overline {q}$. Annihilations to leptons, $W$s, $Z$s,
Higgses and photons have much smaller rates because they are down by
small coupling constants while the annihilation rates to quarks and
gluons are enhanced markedly by large color factors. We estimate the
dominant annihilation processes in perturbation theory. The lowest order
Feynman diagram contributing to the annihilation to quarks is shown in
figure 4a. There is a single gluon exchange. The squared amplitude, after
summing over final state (spin, color and flavor) degeneracies  and
averaging over initial state degeneracies, is
\be \label {ampone}
|M(X \to q\overline {q})|^2 = {{32 \over 9}(4\pi\alpha_{\rm {qcd}})^2 } {|{\vec {p}}|^2
\over p_0^2} \rm {sin}^2\theta \, , 
\ee
where $p$ is the 4 momentum of the $X$ and $\theta$ is the 
scattering angle in the c.m frame. We have taken all the quarks
to be massless. Feynman diagrams corresponding to annihilation to gluons
are shown in figure 4b. The invariant squared amplitude is
\be \label {amptwo}
|M(X \overline {X} \to GG)|^2 = {(4\pi\alpha_{\rm {qcd}})^2\over 9}\left [12
\left ({{\vec {p}}^{\, 2}
\over p_0^2}\right )^2 +6\right ](1+\rm {cos}^2\phi) \, , 
\ee
where $\phi$ is the angle between the two final state gluons in the rest
frame of one of the incoming particles and $p$ is the 4 momentum of the
$X$ in the c.m frame. The large numerical factors in
(\ref {ampone}) and (\ref {amptwo}) come from color and flavor sums. Note that we have
four generations of quarks now.

The thermally averaged cross section can be approximated by the zero
temperature cross section with $|{\vec {p}}| \approx p_0 \approx M_X$. We then
obtain
\be \label {xcross}
\langle \sigma(X \overline {X} \to GG,q\overline {q})\rangle \approx
{(4\pi\alpha_{\rm {qcd}})^2 \over 9\pi M_X^2} \, .
\ee

The $X \overline {X}$ annihilation rate is to be compared with the
baryon number violating decay rates of $X$s. The lowest order Feynman
diagrams for these decays are shown in figure 4c. The squared amplitude
for the decay to two quarks (after summing over final state degeneracies
and averaging over initial state degeneracies) is

\be \label {ampthree}
|M(X \to q_4q_4)|^2 \approx 4 |g|^2 M_{X}^2 \, .
\ee
Once again we have taken the final state particles to be massless. The
corresponding rate for a decay to an anti-quark and an anti-lepton is exactly
the same (larger flavor factor compensates for the smaller color factor):
\be \label {ampfour}
|M(X \to \overline {q_4}\overline {l_4})|^2 \approx 4 |g|^2 M_{X}^2 \, .
\ee

Approximating the thermally averaged decay width by the zero temperature
decay width we get
\be \label {sigmatwo}
\langle \Gamma ({X \rightarrow q_4q_4, \overline {q_4}\overline {l_4}})
\rangle 
\approx {1 \over 2 \pi} |g|^2M_{X} \, .
\ee

Figure 4c shows the Feynman diagram corresponding to the leading order
contribution to the process $q_4q_4 \to \overline {q_4}\overline
{l_4}$. The invariant squared amplitude for a typical process is
\be \label {qqql}
|M(q_4q_4 \to \overline {q_4}\overline{l_4})|^2 \approx 96
|g|^4 {k_0^2 \over M_X^2} \, ,
\ee
where ${k_0}$ is the c.m energy of a $q_4$ in the initial state.
The thermally averaged cross section (summed over all initial and final
state degeneracies)
is approximated by a zero
temperature cross section with $k_0$ set equal to $T_c$. The result is
\be \label {crosql}
\langle \sigma(q_4q_4 \to \overline {q_4}\overline{l_4}) \rangle \approx {12 |g|^4T_c^2 \over
\pi M_X^4} \, .
\ee
Finally the $q_4 \to q_3W$ and $l_4 \to l_3W$ decays (figures 4d) have
the widths
\bear
\langle \Gamma(q_4 \to q_3W)\rangle = {1 \over 8 \pi }{\Theta_{q_4q_3}^2
|g_W|^2 T_c}  \, , \nonumber
\\ \vspace{2mm}
\langle \Gamma(l_4 \to l_3W)\rangle = {1 \over 8 \pi }{\Theta_{l_4l_3}^2
|g_W|^2 T_c} \, , 
\label {qfour}
\eear
where we have made the approximation that the final state
particles are much lighter than the decaying particle and averaged over
initial state degeneracies. We have also approximated the thermal averaging by taking all
masses and momenta in the final expression to be of order
$T_c$. Even with the 
limitations of the above approximations, the expressions in (\ref {qfour})  are
useful as order of magnitude estimates of these decay rates.


\vfil

\end{document}